\documentclass[11pt]{article}
\usepackage[margin=1in]{geometry}
\usepackage{graphicx}
\usepackage{epsfig}
\usepackage{cite}

\def\beq{\begin{equation}}
\def\eeq#1{\label{#1}\end{equation}}
\def\eeqn{\end{equation}}

\def\beqa{\begin{eqnarray}}
\def\eeqa#1{\label{#1}\end{eqnarray}}
\def\eeqan{\end{eqnarray}}

\let\bar=\overbar

\def\Dslash{\not{\hbox{\kern-4pt $D$}}}
\def\dslash{\not{\hbox{\kern-2pt $\del$}}}

\def\msb{{\bar{\ssstyle M \kern -1pt S}}}

\def\Title#1{\begin{center} {\Large {\bf #1} } \end{center}}
\def\Author#1{\begin{center} {\normalsize {\sc #1} } \end{center}}
\def\Institution#1{\begin{center} {\normalsize {\it #1} } \end{center}}
\def\Abstract#1{\noindent {\normalsize {\bf Abstract:} {\normalfont #1}}}
\def\Conference{\vspace{4mm}\begin{raggedright} {\normalsize {\it Talk presented at the 2019 Meeting of the Division of Particles and Fields of the American Physical Society (DPF2019), July 29--August 2, 2019, Boston, C1907293.} } \end{raggedright}\vspace{4mm}}

\def\beq{\begin{equation}}
\def\eeq{\end{equation}}
\def\beqa{\begin{eqnarray}}
\def\eeqa{\end{eqnarray}}

\begin{document}

\Title{Theoretical predictions for top-quark production processes}

\Author{Nikolaos Kidonakis}

\Institution{Department of Physics, Kennesaw State University, Kennesaw, GA 30144, USA}

\Abstract{I present theoretical results through three loops and N$^3$LO for soft-gluon corrections in a variety of processes involving top-quark production. In particular, I present results for total cross sections and differential distributions in single-top production and top-pair production as well as in top-quark processes with electroweak bosons and new physics.}

\Conference

\section{Introduction}

Soft-gluon corrections are theoretically and numerically important for top-quark processes. They are well defined analytically, they approximate known exact results at NLO and NNLO very well, and they generate predictions for even higher-order corrections \cite{NKtop}. Thus, they are important in providing accurate theoretical predictions with a precision that can match the decreasing uncertainties of the experimental collider data. 

We consider a variety of processes involving the production of top-antitop pairs or single top-quarks. For a top-quark process of the general form 
$$ f_{1}(p_1)\, + \, f_{2}\, (p_2) \rightarrow t(p_t)\, + \, X $$
we define the usual variables $s=(p_1+p_2)^2$, $t=(p_1-p_t)^2$, $u=(p_2-p_t)^2$, as well as the threshold variable $s_4=s+t+u-\sum m^2$. As we approach partonic threshold, $s_4 \rightarrow 0$. Soft-gluon corrections contribute terms of the form $[\ln^k(s_4/m_t^2)/s_4]_+$ in the perturbative expansion which can be dominant near partonic threshold.

These soft-gluon corrections can be resummed in double-differential cross sections in Laplace or Mellin moment space. For resummation at next-to-next-to-leading-logarithm (NNLL) accuracy we need two-loop soft anomalous dimensions while at N$^3$LL accuracy we need three-loop soft anomalous dimensions. Finite-order expansions of the resummed cross section are independent of resummation prescriptions, and thus they provide - after matching to known exact results - the most reliable and accurate results at higher orders for cross sections and differential distributions. We denote the results from expansions to second order as approximate NNLO (aNNLO), and from expansions to third order as approximate N$^3$LO (aN$^3$LO).

Soft-gluon resummation is derived in moment space. Taking Laplace 
moments of the partonic cross section with moment variable $N$,
${\hat \sigma}(N)=\int (ds_4/s) \; e^{-N s_4/s} {\hat \sigma}(s_4)$,  
the cross section factorizes in $4-\epsilon$ dimensions as
\beq
\sigma^{f_1 f_2\rightarrow tX}(N,\epsilon)
= H_{IL}^{f_1 f_2\rightarrow tX}\left(\alpha_s(\mu_R)\right) \, 
S_{LI}^{f_1 f_2 \rightarrow tX}\left(\frac{m_t}{N \mu_F},\alpha_s(\mu_R) \right)
\prod  J_{\rm in} \left(N,\mu_F,\epsilon \right)
\prod J_{\rm out} \left(N,\mu_F,\epsilon \right) 
\eeq
where
$H_{IL}^{f_1 f_2\rightarrow tX}$ is a hard function  
and $S_{LI}^{f_1 f_2\rightarrow tX}$ is a soft function, and both are in general matrices in a given color tensor basis.

The soft function satisfies the renormalization group equation
\beq
\left(\mu \frac{\partial}{\partial \mu}
+\beta(g_s)\frac{\partial}{\partial g_s}\right)\,S_{LI}^{f_1 f_2\rightarrow tX}
=-(\Gamma_S^{f_1 f_2\rightarrow tX})_{LK}^{\dagger} S_{KI}^{f_1 f_2\rightarrow tX}
-S_{LK}^{f_1 f_2\rightarrow tX} \Gamma_{S \, KI}^{f_1 f_2\rightarrow tX} \, .
\eeq
The process-dependent soft anomalous dimension matrix, $\Gamma_S^{f_1 f_2\rightarrow tX}$, controls the evolution of the soft function and gives the exponentiation of logarithms of $N$.
For a review of soft-gluon corrections in top-quark processes, see Ref. \cite{NKtop}.

\section{Top-antitop pair production}

\begin{figure}[htb]
\begin{center}
\epsfig{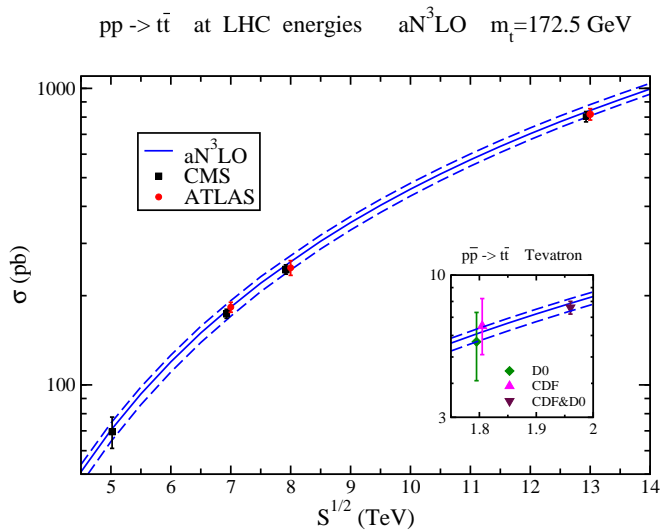}
\caption{The aN$^3$LO top-antitop pair production cross section at LHC energies and (inset) Tevatron energies compared with data from ATLAS \cite{ATLASttlhc}, CMS \cite{CMSttlhc}, CDF \cite{CDF}, D0 \cite{D0} and Tevatron combination \cite{Tevatrontt}.}
\label{ttSaN3LO}
\end{center}
\end{figure} 

At leading order, $t{\bar t}$ production proceeds via two distinct partonic channels, $q{\bar q} \rightarrow t{\bar t}$ and $gg \rightarrow  t{\bar t}$. The color structures of these channels are non-trivial, and matrices are needed to describe the soft anomalous dimension in an appropriate color basis. 

For the $q{\bar q} \rightarrow t{\bar t}$ channel, the soft anomalous dimension is a $2\times 2$ matrix. We use an $s$-channel singlet-octet color tensor basis. At two loops for $q{\bar q} \rightarrow t{\bar t}$, the four matrix elements can be given in terms of the one-loop elements and the cusp anomalous dimension in the form \cite{NKtop,NK3l}
\beqa
\Gamma^{q{\bar q\rightarrow t{\bar t}}\,(2)}_{S\, 11}&=&\Gamma_{\rm cusp}^{(2)} \, , \quad \quad
\Gamma^{q{\bar q}\rightarrow t{\bar t}\,(2)}_{S\, 12}=
\left(K^{'(2)}-C_A N_S^{(2)}\right) \Gamma^{q{\bar q}\rightarrow t{\bar t} \,(1)}_{S\, 12} \, ,
\nonumber \\ 
\Gamma^{q{\bar q}\rightarrow t{\bar t} \,(2)}_{S\, 21}&=&
\left(K^{'(2)}+C_A N_S^{(2)}\right) \Gamma^{q{\bar q}\rightarrow t{\bar t} \,(1)}_{S\, 21} \, ,
\nonumber \\
\Gamma^{q{\bar q}\rightarrow t{\bar t} \,(2)}_{S\, 22}&=& K^{'(2)} \Gamma^{q{\bar q}\rightarrow t{\bar t} \,(1)}_{S\, 22}+\left(1-\frac{C_A}{2C_F}\right)
\left(\Gamma_{\rm cusp}^{(2)}-K^{'(2)}\Gamma_{\rm cusp}^{(1)}\right) \, ,
\eeqa
where $K^{'(2)}=C_A(67/36-\zeta_2/2)-5n_f/18$ and the superscripts indicate the number of loops. For the $gg \rightarrow t{\bar t}$ channel the soft anomalous dimension is a $3\times 3$ matrix (see \cite{NKtop,NK3l} for more details). The structure of the results at three loops is similar, up to four-parton correlations \cite{NK3l}.

We now consider top-antitop pair production at aN$^3$LO with NNLL accuracy \cite{NKtopaN3LO}. For numerical results in this section we use MMHT2014 NNLO pdf \cite{MMHT2014}. In Fig. \ref{ttSaN3LO} we display the theoretical aN$^3$LO cross section (i.e. exact NNLO plus soft-gluon N$^3$LO corrections), together with theoretical uncertainty from scale variation and pdf, at LHC energies and (inset) Tevatron energies, and compare them with recent available data. These are the best theoretical results available since they are based on NNLL resummation of the double-differential cross section without using any resummation prescriptions. The formalism used in our results has consistently been the most successful in postdicting exact NLO results and predicting exact NNLO results for both total cross sections and differential distributions (see Ref. \cite{NKtop} for more discussion and comparisons of various predictions). Furthermore, we observe an excellent agreement of the data with our theoretical predictions at all energies.

\begin{figure}[htb]
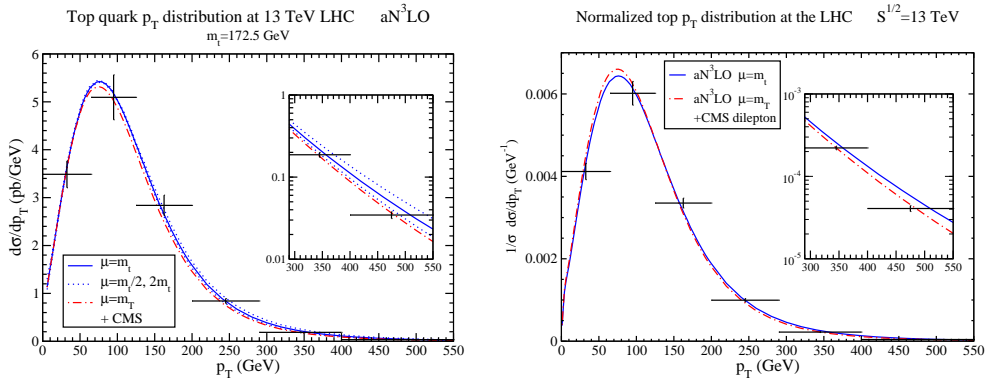

\begin{center}
\epsfig{file=ptaN3LO13lhcCMS1811.06625dileptplot.eps,height=5cm} 
\hspace{2mm}
\epsfig{file=ptnormaN3LO13lhcCMS1811.06625dileptplot.eps,height=5cm} 
\caption{Top-quark $p_T$ distributions (left) and normalized $p_T$ distributions (right) compared with CMS data \cite{CMSpty13dilepton}.}
\label{pTlhc}
\end{center}
\end{figure} 

In Fig. \ref{pTlhc} we show the top-quark $p_T$ distributions, $d\sigma/dp_T$, in $t{\bar t}$ production in the left plot, and the normalized $p_T$ distributions, $(1/\sigma) d\sigma/dp_T$, in the plot on the right, all at 13 TeV energy. We use two choices of central scale, $\mu=m_t$ and $\mu=m_T=(p_T^2+m_t^2)^{1/2}$. We find very good agreement with data from CMS \cite{CMSpty13dilepton}; the latter scale choice works better in describing the data at higher $p_T$.    

\begin{figure}[htb]
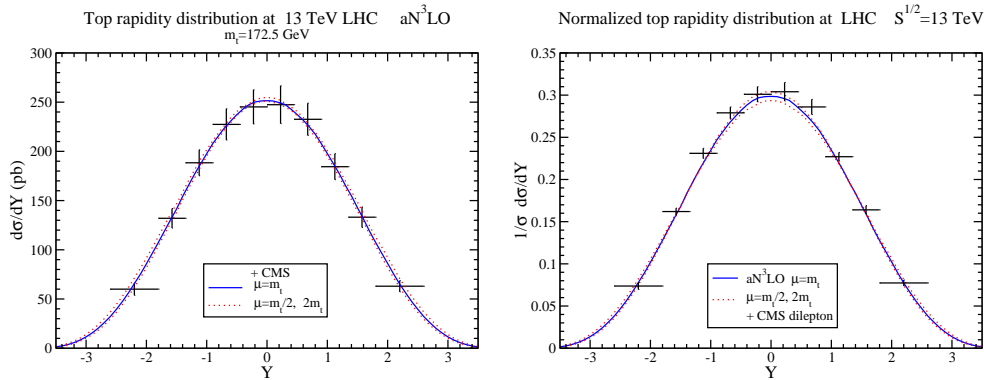

\begin{center}
\epsfig{file=yaN3LO13lhcCMS1811.06625dileptplot.eps,height=5cm} 
\hspace{2mm}
\epsfig{file=ynormaN3LO13lhcCMS1811.06625dileptplot.eps,height=5cm} 
\caption{Top-quark rapidity distributions (left) and normalized rapidity distributions (right) compared with CMS data \cite{CMSpty13dilepton}.}
\label{ylhc}
\end{center}
\end{figure} 

In Fig. \ref{ylhc} we show the top-quark rapidity distributions, $d\sigma/dY$, in the left plot, and the normalized rapidity  distributions, $(1/\sigma) d\sigma/dY$, in the plot on the right, at 13 TeV energy. Again, we find good agreement with data from CMS \cite{CMSpty13dilepton} in both plots.

\begin{figure}[htb]
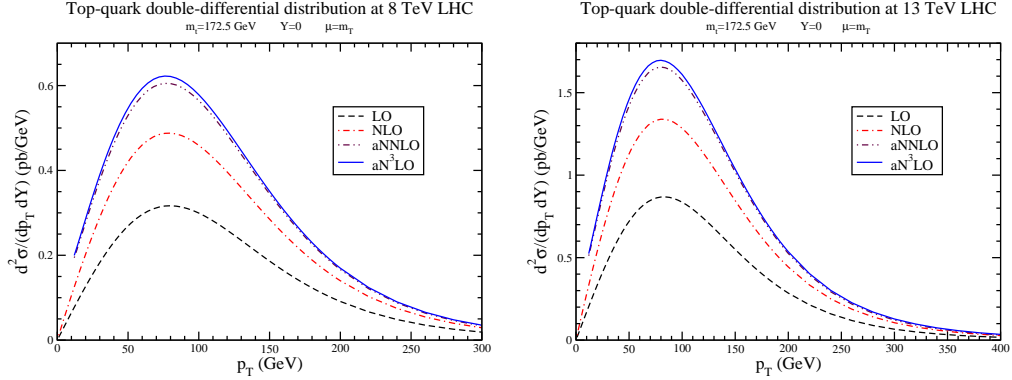

\begin{center}
\epsfig{file=topy0lhc8plot.eps,height=5cm} 
\hspace{2mm}
\epsfig{file=topy0lhc13plot.eps,height=5cm} 
\caption{Top-quark double-differential distributions in $p_T$ and rapidity for central value of the rapidity at (left) 8 TeV and (right) 13 TeV energies.}
\label{doublediff}
\end{center}
\end{figure} 

In Fig. \ref{doublediff} we show select top-quark double-differential distributions in $p_T$ and rapidity, $d^2\sigma/dp_T dY$, at 8 TeV and 13 TeV LHC energies. 

\section{Single-top production}

\begin{figure}[htb]
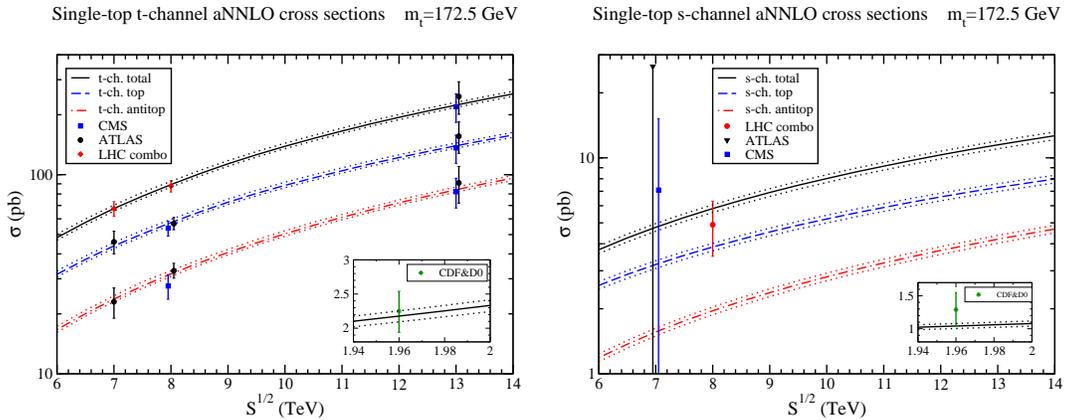

\begin{center}
\epsfig{file=tchplot.eps,height=5.5cm}
\hspace{2mm}
\epsfig{file=schplot.eps,height=5.5cm}
\caption{(Left) aNNLO single-top $t$-channel cross sections compared with data at the LHC \cite{LHC13tch,LHC7and8singletop} and (inset) the Tevatron \cite{CDFD0tch}; (right) aNNLO single-top $s$-channel cross sections compared with data at the LHC \cite{LHC7and8singletop,LHCsch} and (inset) the Tevatron \cite{CDFD0sch}.}
\label{tchsch}
\end{center}
\end{figure}

We continue with a discussion of single-top production for all three Standard Model channels \cite{NKsingletop}. The soft anomalous dimensions are now known for these processes to three loops \cite{NK3loop}. aNNLO results are presented for $t$-channel production and for $s$-channel production while aN$^3$LO results are given for $tW$ production. All numerical results in this section use MMHT2014 NNLO pdf \cite{MMHT2014}.

\subsection{Single-top $t$-channel production}

We begin with single-top $t$-channel production \cite{NKsingletop} via processes $bq \rightarrow tq'$.
The soft anomalous dimension is a $2 \times 2$ matrix. We use a $t$-channel singlet-octet color basis. Results for the soft anomalous dimension matrix are available through three loops \cite{NK3l,NK3loop}.

For the first matrix element we have at one, two, and three loops, respectively \cite{NK3loop}
\beqa
&& \hspace{-5mm} {\Gamma}_{S\, 11}^{bq\rightarrow tq' \, (1)}=
C_F \left[\ln\left(\frac{t(t-m_t^2)}{m_t s^{3/2}}\right)-\frac{1}{2}\right] \, ,
\quad
\Gamma_{S\,11}^{bq\rightarrow tq' \, (2)}= K^{'(2)} \Gamma_{S\,11}^{bq\rightarrow tq' \, (1)}+\frac{1}{4} C_F C_A (1-\zeta_3)\, , 
\nonumber \\ && \hspace{-5mm}
\Gamma_{S\,11}^{bq\rightarrow tq' \, (3)}= K^{'(3)} \Gamma_{S\,11}^{bq\rightarrow tq' \, (1)}+ \frac{1}{2} K^{(2)} C_A (1-\zeta_3)
+C_F C_A^2\left[-\frac{1}{4}+\frac{3}{8}\zeta_2-\frac{\zeta_3}{8}
-\frac{3}{8}\zeta_2 \zeta_3+\frac{9}{16} \zeta_5\right] .
\eeqa
Results for the other matrix elements in $t$-channel production can be found in \cite{NK3loop}.

In the left plot of Fig. \ref{tchsch} we present numerical results for $t$-channel production at aNNLO with NNLL accuracy at LHC and (inset) Tevatron energies. Results are given separately for single-top and single-antitop $t$-channel production, and also for their sum. Very good agreement is observed with data from the LHC and the Tevatron. We also note that NNLO results for $t$-channel production have appeared in Refs. \cite{BCM,BGYZ}.

\subsection{Single-top $s$-channel production}

We continue with single-top $s$-channel production \cite{NKsingletop} via processes $q{\bar q}' \rightarrow t {\bar b}$. The soft anomalous dimension is again a $2 \times 2$ matrix. We use an $s$-channel singlet-octet color basis. Results for the matrix are available through three loops \cite{NK3l,NK3loop}.

For the first matrix element we have  at one, two, and three loops, respectively \cite{NK3loop}
\beqa
&& \hspace{-5mm} \Gamma_{S\, 11}^{q{\bar q}' \rightarrow t {\bar b} \, (1)}=C_F \left[\ln\left(\frac{s-m_t^2}{m_t\sqrt{s}}\right)-\frac{1}{2}\right] \, , \quad 
\Gamma_{S\, 11}^{q{\bar q}' \rightarrow t {\bar b}\, (2)}=K^{'(2)} \Gamma_{S\,11}^{q{\bar q}' \rightarrow t {\bar b} \, (1)}+\frac{1}{4} C_F C_A (1-\zeta_3) \, , 
\nonumber \\ && \hspace{-5mm}
\Gamma_{S\, 11}^{q{\bar q}' \rightarrow t {\bar b} \, (3)}= K^{'(3)} \Gamma_{S\, 11}^{q{\bar q}' \rightarrow t {\bar b} \, (1)}
+\frac{1}{2} K^{(2)} C_A (1-\zeta_3)
+C_F C_A^2\left[-\frac{1}{4}+\frac{3}{8}\zeta_2-\frac{\zeta_3}{8}
-\frac{3}{8}\zeta_2 \zeta_3+\frac{9}{16} \zeta_5\right] .
\eeqa
Results for the other matrix elements in $s$-channel production can be found in \cite{NK3loop}.

In the right plot of Fig. \ref{tchsch} we present numerical results for $s$-channel production at aNNLO with NNLL accuracy at LHC and (inset) Tevatron energies. Results are given separately for single-top and single-antitop $s$-channel production, as well as for their sum. The collider data have large error bars, especially at 7 TeV energy, but good agreement is observed with the theoretical predictions. We also note that NNLO results for $s$-channel production have appeared in Ref. \cite{LG}.

\subsection{Associated $tW$ production}

\begin{figure}[htb]
\begin{center}
\epsfig{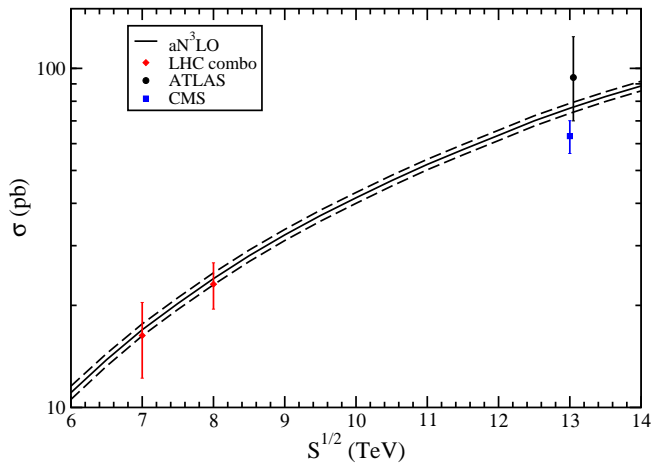}
\caption{aN$^3$LO single-top $tW$ cross sections compared with data at the LHC \cite{LHC7and8singletop,LHC13tW}.}
\label{tW}
\end{center}
\end{figure}

Single-top production can also proceed via associated production with a $W$ boson, $bg \rightarrow tW^-$ \cite{NKsingletop}. In this case the soft anomalous dimension is a simple function and it is given 
at one loop by
\beq
\Gamma_S^{bg \rightarrow tW \, (1)}=C_F \left[\ln\left(\frac{m_t^2-t}{m_t\sqrt{s}}\right)-\frac{1}{2}\right] +\frac{C_A}{2} \ln\left(\frac{u-m_t^2}{t-m_t^2}\right) \, ,
\eeq
at two loops by
\beq
\Gamma_S^{bg \rightarrow tW (2)}=K^{'(2)} \Gamma_S^{bg \rightarrow tW \, (1)}
+\frac{1}{4}C_F C_A (1-\zeta_3) \, ,
\eeq
and at three loops by \cite{NK3loop}
\beq
\Gamma_S^{bg \rightarrow tW \, (3)}=K^{'(3)} \Gamma_S^{bg \rightarrow tW \, (1)}+\frac{1}{2} K^{(2)} C_A (1-\zeta_3)+C_F C_A^2\left[-\frac{1}{4}+\frac{3}{8}\zeta_2-\frac{\zeta_3}{8}-\frac{3}{8}\zeta_2 \zeta_3+\frac{9}{16} \zeta_5\right]\, .
\eeq

In Fig. \ref{tW} we show theoretical predictions for $tW$ production at aN$^3$LO with NNLL accuracy at LHC energies. Again, we oberve very good agreement with LHC data. We also note that the numerical effect of adding terms involving $\Gamma_S^{tW \, (3)}$ (contributing to N$^3$LL accuracy) is very small.

\section{$t Z'$ production}

\begin{figure}[htb]
\begin{center}
\epsfig{file=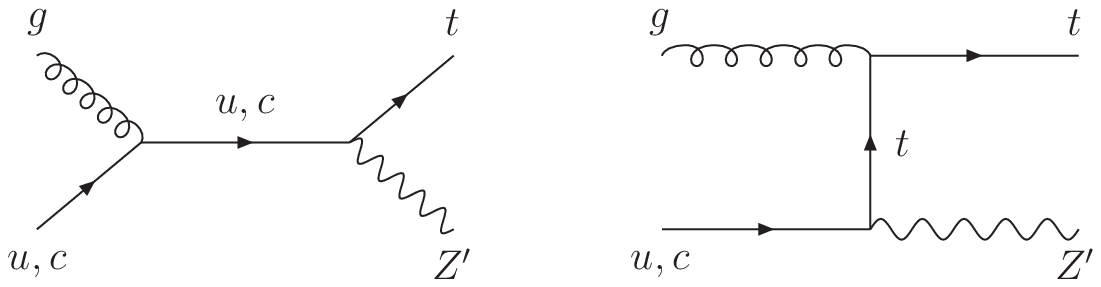,height=23mm}
\hspace{5mm}
\epsfig{file=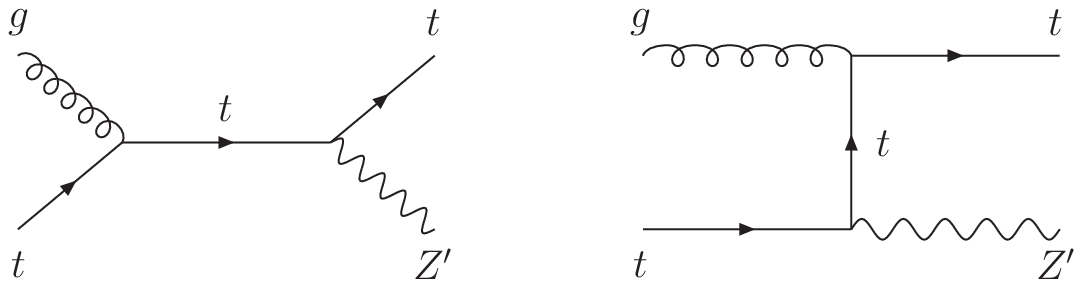,height=23mm}
\caption{Leading-order diagrams for $tZ'$ production via top-quark anomalous couplings, $gq \rightarrow tZ'$ (left two diagrams), and via top-initiated processes, $gt \rightarrow tZ'$ (right two diagrams).}
\label{tZpdiag}
\end{center}
\end{figure} 

Finally, we consider $tZ'$ production \cite{tZp} via top-quark anomalous couplings, $gq \rightarrow tZ'$ (left two diagrams in Fig. \ref{tZpdiag}), and with initial-state top quarks, $gt \rightarrow tZ'$ (right two diagrams in Fig. \ref{tZpdiag}).

\begin{figure}[htb]
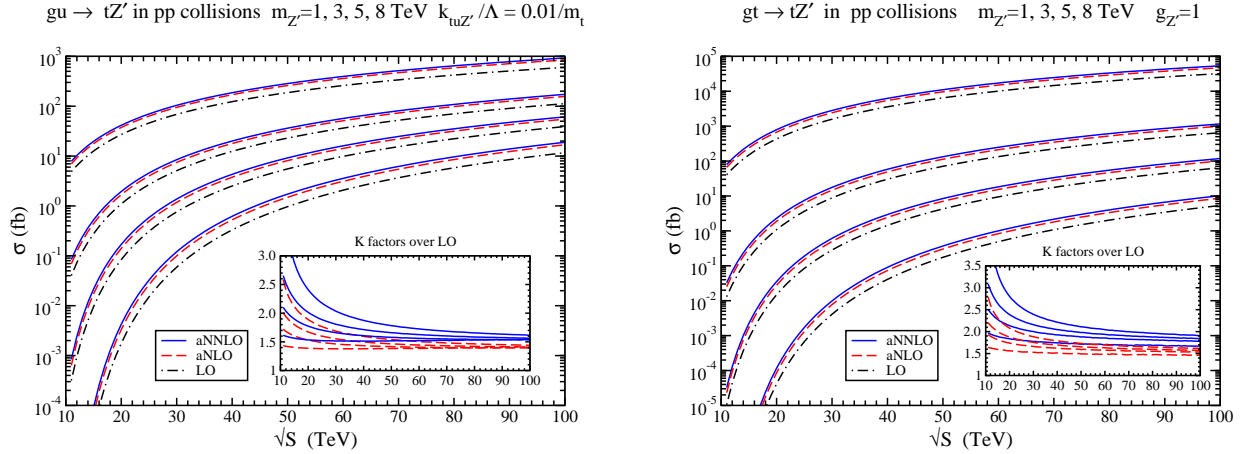

\begin{center}
\epsfig{file=gutZprootSplot.eps,height=6cm}
\hspace{8mm}
\epsfig{file=tZprootSplot.eps,height=6cm}
\caption{Cross sections for (left) $gu \rightarrow tZ'$ and (right) $gt \rightarrow tZ'$ versus collider energy for four choices of $Z'$ mass. The inset plots show $K$-factors relative to LO.}
\label{tZp}
\end{center}
\end{figure} 

Numerical results for four different values of $Z'$ mass are shown for the $gu \rightarrow tZ'$ cross section as a function of collider energy in the left plot of Fig. \ref{tZp} using CT14 NNLO pdf \cite{CT14}, and for the $gt \rightarrow tZ'$ cross section in the right plot of Fig. \ref{tZp} using NNPDF3.1 NNLO pdf \cite{NNPDF3.1}.

Related results for $tZ$ and $t\gamma$ production via top-quark anomalous couplings have been presented in \cite{tZ,tgam}, and for $tH^-$ production in \cite{NKtH}.

\section{Summary}

We have studied soft-gluon corrections for top-quark processes through three loops. We presented results for $t{\bar t}$ production at aN$^3$LO, 
$t$-channel and $s$-channel single-top production at aNNLO, 
and $tW$ production at aN$^3$LO.
Excellent agreement is found between theoretical predictions and collider data at the LHC and the Tevatron.
We also showed results for $tZ'$ production through aNNLO in various models of new physics. Similar results apply to $tZ$, $t\gamma$, and $tH^-$ production at aNNLO and aN$^3$LO. Higher-order soft-gluon corrections are very significant in all cases.

\section*{Acknowledgements}
This material is based upon work supported by the National Science Foundation 
under Grant No. PHY 1820795.

\end{document}